\documentclass[conference]{IEEEtran}
\IEEEoverridecommandlockouts  
\usepackage{cite}

\usepackage{placeins}

\usepackage{soul}

\usepackage{float}

\usepackage{url}

\ifCLASSINFOpdf
\else
\fi
\usepackage{amsmath}
\usepackage{graphicx}
\usepackage{subcaption}

\usepackage{url}
\usepackage{graphicx} 

\usepackage{amssymb}

\usepackage{float} 
\usepackage{titlesec} 
\usepackage{xcolor}

\begin{document}
%
\title{Fusion of Cellular ISAC and Passive RF Sensing for UAV Detection and Tracking%
\thanks{This work is supported in part by the NSF awards CNS-1939334, CNS-2332835, and DGE-2137100.}
}

\author{%
    \IEEEauthorblockN{
        Cole Dickerson, Sean Kearney, Sultan Manjur, Ismail G\"{u}ven\c{c},
        Sevgi Gurbuz, Ali Gurbuz, \\ \"{O}zg\"{u}r \"{O}zdemir, Mihail Sichitiu
    }
    \IEEEauthorblockA{Department of Electrical and Computer Engineering,\\
    North Carolina State University, Raleigh, NC, USA}
    \IEEEauthorblockA{Email: \{jcdicker, sjkearne, smanjur, iguvenc, szgurbuz, acgurbuz, oozdemi, mlsichit\}@ncsu.edu}
}


%


\maketitle
\begin{abstract}
The rapid growth of unmanned aerial vehicles (UAVs) in civilian and critical-infrastructure airspace has created a need for reliable detection and tracking systems that operate under diverse environmental and sensing conditions. This paper presents a UAV detection and tracking system that fuses measurements from a network of passive Keysight N6841A RF sensors and a Ku-band Fortem TrueView R20 radar operating in the FR3 spectrum (16.3 GHz) as an ISAC proxy. Real-world experiments at the NSF AERPAW testbed demonstrate that radar and RF sensing provide complementary strengths under varying geometric, range, and line-of-sight conditions. A Kalman filter using a constant-velocity motion model integrates the asynchronous 2D RF and 3D radar observations, suppressing large standalone errors, improving accuracy over individual modalities, and increasing tracking coverage without degrading performance. These results demonstrate the effectiveness of multi-modal, ISAC-oriented sensing for robust UAV tracking in outdoor environments.
\end{abstract}


%
\IEEEpeerreviewmaketitle

\section{Introduction}

The decreasing cost and widespread availability of unmanned aerial vehicles (UAVs) have accelerated their use across defense, agriculture, communications, public safety, and spectrum monitoring applications. Their growing presence also raises security and airspace management challenges around critical infrastructure, motivating reliable UAV detection and tracking for unmanned traffic management (UTM) \cite{faa2020utm, clemente2021radar}. Recent work has investigated RF- and radar-based sensing for non-cooperative UAVs that withhold telemetry or identification signals \cite{clemente2021radar, 10938573}, underscoring the need for low-cost, infrastructure-based solutions that do not depend on UAV cooperation. Emerging 5G/6G systems further enable integrated sensing and communication (ISAC), potentially transforming cellular networks into distributed sensing platforms.

\begin{figure}
    \centering
    \includegraphics[width=1\linewidth]{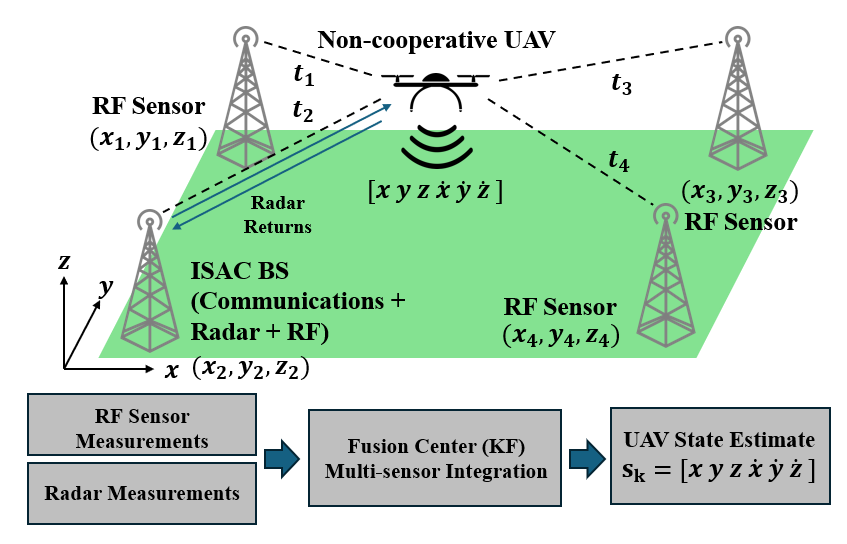}
    \caption{UAV tracking setup showing a non-cooperative UAV, three RF sensors, and an ISAC base station (BS) providing communications, radar, and RF sensing. TDOA and radar measurements are fused for UAV state estimation.}
    \label{fig:sysmodel}
\end{figure}

\begin{figure*}[t]
    \centering
    \begin{subfigure}[t]{0.32\textwidth}
        \centering
        \includegraphics[width=\linewidth]{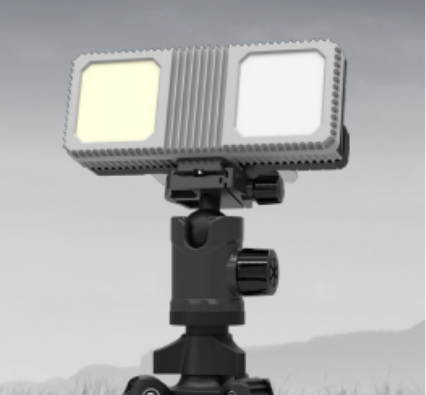} 
        \caption{Fortem TrueView R20 radar.}
        \label{fig:fortem_radar}
    \end{subfigure}
    \hfill
    \begin{subfigure}[t]{0.32\textwidth}
        \centering
        \includegraphics[width=\linewidth]{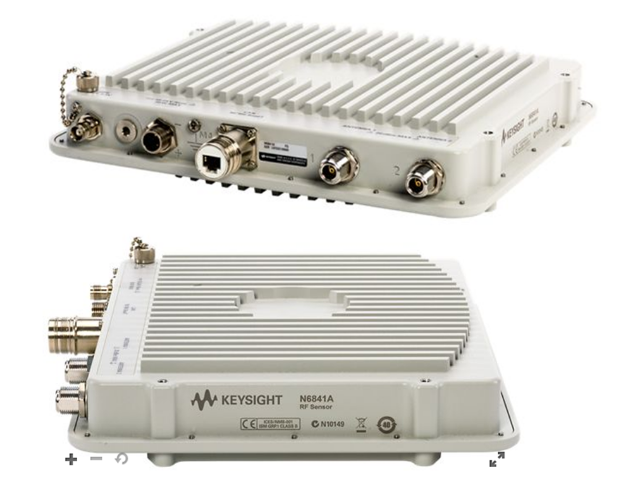} 
        \caption{Keysight N6841A RF sensor.}
        \label{fig:rf_sensor}
    \end{subfigure}
    \hfill
    \begin{subfigure}[t]{0.32\textwidth}
        \centering
        \includegraphics[width=\linewidth]{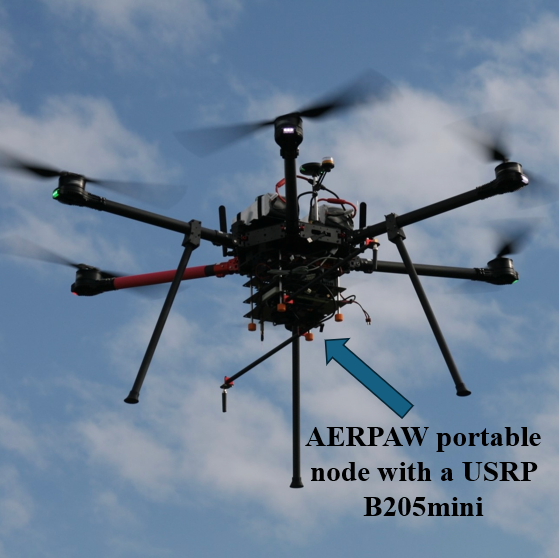}
        \caption{AERPAW UAV with portable USRP node.}
        \label{fig:uav}
    \end{subfigure}

    \caption{Sensing hardware and UAV platform used in the experiment at AERPAW's LWFRL site. }
    \label{fig:three_panel_sensors_uav}
\end{figure*}

\begin{figure}
    \centering
    \includegraphics[width=1\linewidth]{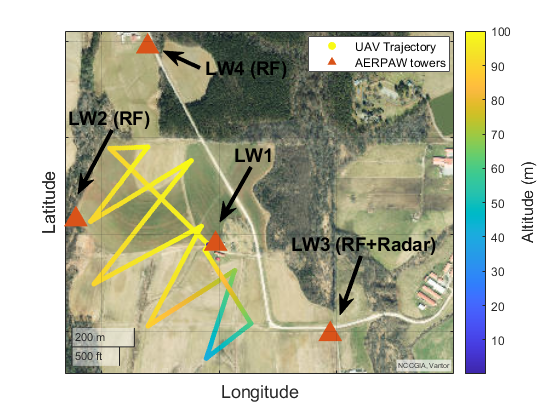}
    \caption{UAV trajectory at the LWFRL, colored by altitude.}
    \label{fig:gttraj}
\end{figure}

Radar and RF sensing offer complementary strengths for UAV detection and tracking \cite{8337900}. Radars provide accurate range and Doppler measurements and do not rely on UAV transmissions, but they struggle at longer ranges with small radar cross-section (RCS) targets. RF sensors are low cost, covert, and can function in non-line-of-sight (NLOS) scenarios, but depend on favorable geometry and the presence of RF emissions. Fusing these modalities is therefore essential for achieving robust, infrastructure-based UAV tracking across diverse conditions.

Motivated by these sensing challenges, recent research has investigated a wide range of approaches for UAV detection and tracking across radar, RF, and emerging ISAC systems. Studies have shown that 5G and 6G ISAC systems can repurpose cellular signals for UAV detection and localization \cite{5GNRSSB, realworldISAC}. RF sensor networks using time-difference-of-arrival (TDOA) localization have also been shown to reliably track non-cooperative UAVs in outdoor settings \cite{ICCDickerson, bhattacherjee2022experimental}.

To further support research on UAV tracking and sensor fusion, this paper makes two key contributions. First, we develop a Kalman filter with a constant-velocity motion model that fuses radar and RF sensor position estimates to enable robust UAV tracking in outdoor environments. This fusion framework leverages the complementary strengths of TDOA RF sensing and radar UAV tracking, improving performance over stand-alone sensing modalities. Second, we release a comprehensive public dataset containing multiple UAV ground-truth trajectories, radar measurements, and RF sensor data, providing a valuable resource for benchmarking UAV sensing, tracking, and sensor-fusion algorithms \cite{AERPAW_UAV_RFS_Radar_2025}.

\section{Experimental Setup and Measurement}

The experiments in this work were conducted at the Aerial Experimentation and Research Platform for Advanced Wireless (AERPAW), the first large-scale wireless research testbed that supports research at the convergence of 5G technologies and autonomous drones. AERPAW integrates UAVs, RF sensing nodes, and a radar unit within a unified test environment, enabling precise repeatable flight trajectories alongside RF sensor and radar data collection. The measurements used in this study were collected at the Lake Wheeler Field Road Labs (LWFRL), one of AERPAW's outdoor test sites. Fig.~\ref{fig:gttraj} shows the UAV's altitude-colored ground-truth trajectory and the locations of four AERPAW towers (LW1-LW4). A comprehensive overview of the AERPAW platform, sensing capabilites, and previously released datasets is provided in \cite{raouf2025wirelessdatasets}.

The RF sensing nodes used in this experiment are mounted on fixed tower sites across the field. Towers LW1–LW4 lie within the primary flight area shown in Fig.~\ref{fig:gttraj}, while tower LW5 is positioned approximately 1.1 km north of LW4 and therefore lies outside the region shown. Towers LW2-5 each host a Keysight N6841A RF sensor, and the Fortem R20 radar was deployed on a 10 m mast adjacent to tower LW3, providing co-located RF and radar sensing at that location.

\subsection{UAV Platform and Ground Truth}

A custom AERPAW hexacopter equipped with an onboard Intel NUC-10 (i7-10710U, 64 GB RAM) served as the mobile platform for data collection in this study (Fig.~\ref{fig:uav}). The UAV carried a Real-Time Kinematic (RTK)-enabled GNSS receiver that provided high-accuracy position logs during each flight. In addition, a Universal Software Radio Peripheral (USRP) B210 software-defined radio mounted on the UAV transmitted a 5 MHz channel-sounding waveform centered at 3.32 GHz. Rather than relying on control signals or telemetry, this waveform acted as a non-cooperative RF emission that the network of RF sensors tracked to enable TDOA localization.

\subsection{RF Sensor Nodes (Keysight N6841A)}

Keysight N6841A RF sensors were used to capture the UAV's transmitted waveform and extract TDOA measurements (Fig.~\ref{fig:rf_sensor}). Each unit operates from 20 MHz to 6 GHz with up to 20 MHz of instantaneous bandwidth and includes a broadband omnidirectional antenna with GPS-disciplined timing for precise signal timestamping and cross-correlation. Through the N6854A software suite, the sensors support TDOA, received signal strength (RSS), and hybrid geolocation of RF emitters within roughly a 2 km radius. In AERPAW, the N6841A units are synchronized through a shared GPS timing infrastructure and managed on a common subnet via the Keysight Geolocation Server.

\subsection{Radar System (Fortem TrueView R20)}

The Fortem TrueView R20 is a Ku-band electronically scanned radar used  to provide position estimates for UAV detection and tracking (Fig.~\ref{fig:fortem_radar}). The radar offers meter-level range resolution, wide field-of-view (FOV) coverage, and real-time onboard tracking. The key operating parameters of the R20 relevant to our experimental setup are summarized in Table~\ref{tab:r20specs}.

\begin{table}[t]
\centering
\caption{Specifications of the Fortem TrueView R20 radar.}
\label{tab:r20specs}
\begin{tabular}{l c}
\hline
\textbf{Parameter} & \textbf{Value} \\
\hline
Frequency Range & 15.4--16.7 GHz \\
Maximum Bandwidth & 180 MHz (1 m range resolution) \\
Tx Antenna Gain & 12 dBi \\
Tx EIRP & 30 W (+44.8 dBm) \\
Field of View & 120$^{\circ}$ az., 60$^{\circ}$ el. \\
Angular Accuracy & $\pm 2^{\circ}$ az., $\pm 2^{\circ}$ el. \\
Track Update Rate & 64 ms -- 1.3 s \\
Minimum Radial Velocity & 0.15 m/s \\
Number of RX Channels & 4 \\
\hline
\end{tabular}
\end{table}

Although the R20 is a commercial Ku-band radar, its operating frequency falls squarely within the FR3 (7-24 GHz) band that is now a key target for 6G ISAC research. FR3 is considered well-suited for ISAC because it offers a favorable trade-off between mmWave-like sensing resolution and sub-6 GHz coverage, and because existing X/Ku/K-band radars already provide mature hardware and algorithmic foundations that can be adapted for ISAC operation \cite{FR3}. Using the R20 therefore provides a realistic proxy for an FR3-band ISAC sensor, particularly with respect to range-angle resolution and outdoor performance, making it an appropriate stand-in for evaluating ISAC UAV tracking.

\subsection{Preprocessing of RF Sensor and Radar Measurements}

The RF sensors output 2D latitude and longitude, while the radar provides 3D latitude, longitude, and altitude. All measurements were transformed into a common local East-North-Up (ENU) frame using MATLAB's \textit{geodetic2enu} function with LW1 as the origin, ensuring a consistent Cartesian coordinate system for fusion. Ground-truth positions were time-aligned to the sensor measurements using linear interpolation.

\section{Kalman Filter with Constant-Velocity Motion Model for Sensor Fusion}

With the sensing infrastructure and measurement modalities established, we next describe the fusion framework used to combine radar and RF sensor measurements into a unified UAV state estimate. Because both sensing systems provide asynchronous and noisy position updates, we employ a Kalman filter (KF) with a constant-velocity (CV) motion model (MM) to integrate their complementary strengths and generate a smooth trajectory estimate.

\subsection{Motion and Measurement Models}

Discrete-time KFs estimate the current state of a target by combining past state information and newly available sensor measurements. A tracking problem is typically formulated using a dynamic MM and a corresponding observation model, expressed as:
\begin{equation}
\begin{aligned}
\textbf{State model:} \quad & \mathbf{s}_k = f_k(\mathbf{s}_{k-1}) + \mathbf{w}_{k-1}, \\
\textbf{Observation model:} \quad & \mathbf{z}_k = \mathbf{H} \mathbf{s}_k + \mathbf{v}_k ,
\end{aligned}
\end{equation}
where $k$ denotes the discrete time step index $k \in\{1, \ldots, K\}$. The vector $\mathbf{s}_k$ represents the UAV's state at time step $k$, $f_k(\cdot)$ is the MM function describing its evolution, and $\mathbf{w}_k$ is the zero-mean Gaussian process noise with covariance $\mathbf{Q}_k$. The measurement $\mathbf{z}_k$ is related to the underlying state through the observation matrix $\mathbf{H}$, and $\mathbf{v}_k$ is zero-mean Gaussian measurement noise with covariance $\mathbf{R}_k$. 

The UAV dynamics are modeled using a CV MM, which assumes the UAV maintains a fixed velocity between measurement steps. This assumption matches the AERPAW waypoint flight profiles, where the UAV travels along straight-line segments at nearly uniform speed for most of the flight. The state vector $\mathbf{s}_k$ and transition matrix $\mathbf{F}_k$ are expressed as:
\begin{equation}
\setlength{\arraycolsep}{1.5pt}
\mathbf{s}_k^{\mathrm{CV}}
=
\begin{bmatrix}
x\; & y\; & z\; & \dot{x}\; & \dot{y}\; & \dot{z}\;
\end{bmatrix}^\mathrm{T},
\qquad
\mathbf{F}_k^{\mathrm{CV}}
=
\begin{bmatrix}
\mathbf{I}_3 & T_k \mathbf{I}_3 \\
\mathbf{0}_3 & \mathbf{I}_3
\end{bmatrix},
\label{eq:cvmodel}
\end{equation}
where the terms $x,y,z$ denote the UAV's 3D ENU position, $\dot{x}, \dot{y}, \dot{z}$ are the corresponding velocity components, and $T_k$ is the time between successive samples used to propagate the state forward.

Because the state vector contains both position and velocity components, while we only use position measurements from our sensors, the KF estimates the velocities from the dynamics. The Fortem radar outputs full 3D position ($x,y,z$), whereas the Keysight RF sensors only provide 2D horizontal position ($x,y$). This difference is captured by the measurement matrices $\mathbf{H}$:
\begin{equation}
\mathbf{H}_\text{rad} = \begin{bmatrix} \mathbf{I}_3 & \mathbf{0}_{3\times 3} \end{bmatrix},
\qquad
\mathbf{H}_\text{RF}  = \begin{bmatrix} \mathbf{I}_2 & \mathbf{0}_{2\times 4} \end{bmatrix},
\label{eq:Hmodels}
\end{equation}
which select either the 2D or 3D position components from the 6-dimensional CV state. A unified expression for both sensing modalities is $\mathbf{z}_{k,i} = \mathbf{H}_i \mathbf{s}_k + \mathbf{v}_{k,i}$, 
$i \in \{\mathrm{rad}, \mathrm{RF}\}$. This notation compactly expresses that both radar and RF sensors observe different subsets of the same underlying UAV state, each with its own measurement dimensionality and noise characteristics.

\subsection{Kalman Filter Update Cycle}

The KF operates in two stages: prediction and update. In the prediction step, the filter propagates the previous state estimate $\hat{\mathbf{s}}_{k-1}$ and covariance $\hat{\mathbf{P}}_{k-1}$ forward in time using the MM, producing the a priori estimates $\hat{\mathbf{s}}_k^{-}$and $\hat{\mathbf{P}}_k^{-}$. When a new radar or RF measurement becomes available, the update step corrects these predictions. The Kalman gain $\mathbf{K}_k$ determines how much the filter trusts the measurement relative to the prediction, yielding the updated a posteriori state estimate $\hat{\mathbf{s}}_k$ and covariance $\hat{\mathbf{P}}_k$. These steps are summarized below:
\begin{equation}
\begin{aligned}
\textbf{Prediction:} \qquad
&\hat{\mathbf{s}}_{k}^{-} = f_k(\hat{\mathbf{s}}_{k-1}), \\
&\hat{\mathbf{P}}_{k}^{-}
    = \mathbf{F}_{k}\, \hat{\mathbf{P}}_{k-1}\, \mathbf{F}_{k}^{\mathrm{T}}
      + \mathbf{Q}_{k}, \\[6pt]
\textbf{Update:} \qquad
&\mathbf{K}_{k} = 
    \hat{\mathbf{P}}_{k}^{-}\mathbf{H}_{i}^{\mathrm{T}}
    \left(\mathbf{H}_{i}\hat{\mathbf{P}}_{k}^{-}\mathbf{H}_{i}^{\mathrm{T}} 
    + \mathbf{R}_{i}\right)^{-1}, \\
&\hat{\mathbf{s}}_{k} = \hat{\mathbf{s}}_{k}^{-} 
    + \mathbf{K}_{k}\!\left(\mathbf{z}_{k,i} - \mathbf{H}_{i}\hat{\mathbf{s}}_{k}^{-}\right), \\
&\hat{\mathbf{P}}_{k} = 
    \left(\mathbf{I} - \mathbf{K}_{k}\mathbf{H}_{i}\right)\hat{\mathbf{P}}_{k}^{-}.
\end{aligned}
\label{eq:kf_update}
\end{equation}

\subsection{Covariance Modeling and Tuning}
The process noise covariance $\mathbf{Q}_k$ follows the standard discrete-time white-acceleration CV model, assuming independent acceleration noise along each axis, using the well-known formulation in \cite[Eq. 6.2.2-12]{barshalom2001estimation}. The measurement noise covariance $\mathbf{R_i}$ for each sensing modality was estimated empirically. Measurement errors were computed relative to the UAV ground-truth, and MATLAB's \textit{cov} function was used to obtain modality-specific covariances. Because RF sensors measure only horizontal $(x,y)$ position, $\mathbf{R}_{\text{RF}}$ is $2 \text{x}2$, while radar provides full $(x,y,z)$ position and yields a $3\text{x}3$ matrix $\mathbf{R}_{\text{rad}}$. These data-driven covariances allow the KF to weight each sensor according to its accuracy in the AERPAW environment.

\subsection{Implementation Details and Measurement Validation}
\label{sec:gate}

The CV Kalman filter processes asynchronous radar and RF measurements in timestamp order. Because both sensing modalities can occasionally produce outliers, each measurement is validated before being used in an update. For each measurement at time $t_k$, the filter performs a prediction using $\Delta t = t_k - t_{k-1}$, selects the appropriate measurement model $\mathbf{H_i}$ and covariance $\mathbf{R_i}$, and applies two gating steps. Radar measurements are first range-gated, discarding detections beyond 800 m, which corresponds to the distance where the R20's errors sharply increase (see Fig.~\ref{fig:radar_error_vs_range}).  All remaining measurements from the radar and RF sensors are then subjected to a normalized innovation squared (NIS) test \cite{4085883}:
\begin{equation}
\epsilon_{k,i}
    = \tilde{\mathbf{y}}_{k,i}^{\mathrm{T}}
      \mathbf{S}_{k,i}^{-1}
      \tilde{\mathbf{y}}_{k,i},
\label{eq:NIS}
\end{equation}
where $\tilde{\mathbf{y}}_{k,i} = z_{k,i} - \mathbf{H}_i\hat{\mathbf{s}}_{k}^{-}$ is the innovation and $\mathbf{S_{k,i}} = \mathbf{H}_i \hat{\mathbf{P}}_k^{-} \mathbf{H}_i^{\mathrm{T}}+\mathbf{R}_i$ its covariance, computed as in the KF update equations. A measurement is accepted only if $\epsilon_{k,i}$ lies within the 95\% chi-squared confidence region for its dimension, ensuring that only statistically consistent observations are fused into the filter. Valid measurements passing both gates trigger a KF update, while rejected measurements are skipped to prevent corrupting the trajectory. When a measurement fails validation, the KF coasts forward using only its prediction step until the next valid radar or RF update. The filter was efficiently implemented in MATLAB, and the six-state CV model supports real-time execution.

\section{Experimental Results and Discussion}
\label{sec:results}

This section presents the experimental evaluation of the radar, RF, and fused tracking system. We begin by examining the behavior of each sensing modality independently to understand how measurement geometry and signal conditions influence performance before analyzing the fused estimates produced by the KF. 

\subsection{Standalone Sensor Performance}

\begin{figure}
    \centering
    \includegraphics[width=1\linewidth]{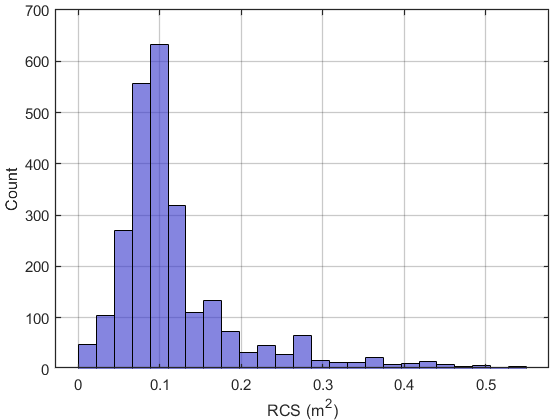}
    \caption{Histogram of measured UAV RCS values ($\text{m}^2$) collected by the Fortem R20 radar.}
    \label{fig:rcs_hist}
\end{figure}

The Fortem radar demonstrated strong standalone performance, maintaining continuous 3D tracking for nearly the entire UAV trajectory. Before validation, the radar produced 3106 measurements with 100\% temporal coverage and a stable 0.25 s sampling interval, indicating reliable line-of-sight (LOS) tracking throughout the flight. Raw errors were relatively low, with a mean of 26.2 m and a standard deviation of 25.5 m. The measured UAV RCS ranged from 0.0066 $\text{m}^2$ to 0.549 $\text{m}^2$, with a mean of 0.12 $\text{m}^2$, consistent with typical multirotor UAVs signatures at Ku-band. Fig.~\ref{fig:rcs_hist} presents a histogram of these RCS measurements, showing a strong peak near typical multirotor values and a long tail of higher-RCS returns caused by aspect-dependent specular reflections.

\begin{figure}[t]
    \centering
    \includegraphics[width=\columnwidth]{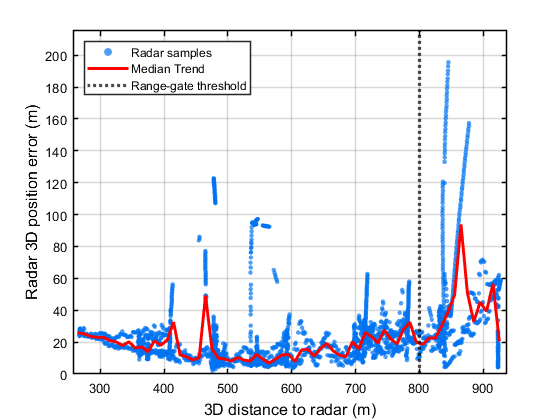}
    \caption{Radar localization error versus 3D distance from radar.}
    \label{fig:radar_error_vs_range}
\end{figure}

\begin{figure}[t]
    \centering
    \includegraphics[width=\columnwidth]{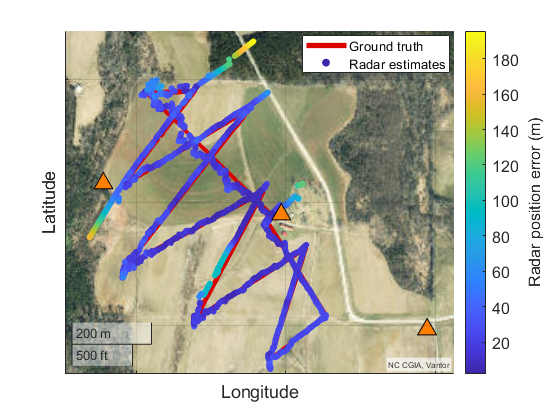}
    \caption{Radar position estimates, color-coded by localization error, over the ground-truth trajectory.}
    \label{fig:radar_traj_error}
\end{figure}

Radar performance degraded beyond roughly 800 m (Fig.~\ref{fig:radar_error_vs_range}), where detections became noisier and occasionally fragmented into smaller tracks. Although Fortem's internal processing is proprietary, the observed behavior suggests an onboard tracking filter that sometimes overcommitted to its predicted trajectory, particularly during sharper turns or decelerations - producing the high-error yellow regions ($>$150 m) in Fig.~\ref{fig:radar_traj_error}. To mitigate these effects, we retained only the largest continuous track and applied range gating and NIS testing as in Section~\ref{sec:gate}, reducing the dataset to 2403 measurements with improved accuracy (mean error of 21.0 m and a standard deviation of 17.1 m). This filtering step, however, reduced the covered portion of the UAV's trajectory by 22.7\%, necessitating an alternative tracking modality. In future experiments, we will restrict the radar's maximum detection range to avoid long-range instabilities and support real-time fusion.

The Keysight RF sensor network also provided extensive standalone coverage of the UAV's trajectory. Prior to validation, the RF sensors produced 206 position estimates out of 216 attempts with 95.4\% temporal coverage. The average sampling interval was 3.88 s, much slower than the radar's 0.25 s rate, making the RF solution more sensitive to geometry changes and transient NLOS conditions. As expected for TDOA-based localization in mixed line-of-sight (LOS) environments \cite{ICCDickerson}, the raw estimates showed large variability, with errors ranging from 2.4 m to 4831.6 m and a mean error of 471.8 m with a standard deviation of 885.2 m before NIS testing.

\begin{figure}[t]
    \centering
    \includegraphics[width=\columnwidth]{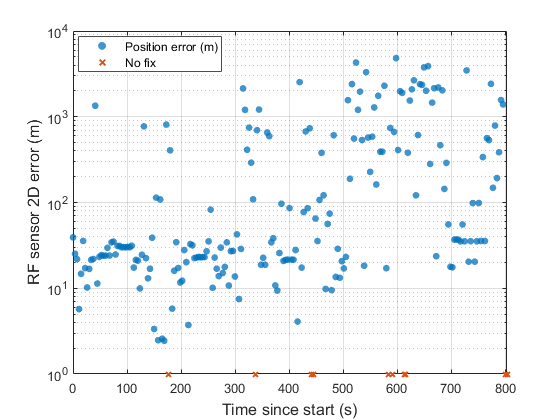}
    \caption{RF sensor localization error versus time since the start of the flight.}
    \label{fig:rf_error_vs_time}
\end{figure}

After applying NIS testing, the RF dataset reduced to 125 valid measurements, but accuracy improved substantially, with a mean error of 25.8 m, a standard deviation of 16.2 m, and a reduced maximum error of 113.9 m. Crucially, the gated results demonstrate a strong complementarity with the radar's performance. As shown in Fig.~\ref{fig:rf_error_vs_time}, the RF sensors performed best in the first half of the trajectory, where the radar struggled most, as long-range radar measurements were removed. This follows naturally from the experimental geometry: the RF sensors were distributed around the field, whereas the radar was mounted near LW3 in the southwest corner. When the UAV traveled north, the RF sensors benefited from strong LOS and favorable multilateration geometry, resulting in their best performance. In contrast, the radar performed best when the UAV was closer to its position near LW3, typically when flying south, where the shorter range and boresight alignment improved measurement accuracy.

\begin{table*}[!t]
\centering
\caption{Position error and track statistics before validation, after validation, and after sensor fusion.}
\small
\begin{tabular}{lcccccc}
\hline
\textbf{Type} & \textbf{Count} & \textbf{Min (m)} & \textbf{Max (m)} & \textbf{Mean (m)} & \textbf{Std (m)} & \textbf{Coverage (\%)} \\
\hline
\multicolumn{7}{l}{\textbf{Before Fusion (Raw Measurements)}} \\
RF Sensor (2D) & 206 & 2.4 & 4831.6 & 471.8 & 885.2 & 95.4 \\
Radar (3D) & 3106 & 2.1 & 195.6 & 26.2 & 25.5 & 100 \\
\hline
\multicolumn{7}{l}{\textbf{After Sensor Validation (NIS Testing, Range Gating, Selection of Largest Track)}} \\
RF Sensor (2D) & 125 & 2.4 & 113.9 & 25.8 & 16.2 & 57.9 \\
Radar (Longest Track, 3D) & 2403 & 2.1 & 97.2 & 21.0 & 17.1 & 77.3 \\
\hline
\multicolumn{7}{l}{\textbf{After Fusion (Kalman Filter Output)}} \\
KF Fused Track (All Steps, 3D) & 2655 & 2.0 & 109.1 & 21.9 & 17.9 & 100 \\
KF Updated Estimates (3D) & 2528 & 2.0 & 109.0 & 21.6 & 17.8 & 95.2 \\
KF Coasted Estimates (3D) & 127 & 4.7 & 109.1 & 28.3 & 18.4 & 4.8 \\
\hline
\end{tabular}
\label{tab:position_error_stats}
\end{table*}

These results show that although the standalone RF solution is more vulnerable to outliers and its slower 3.88 s update rate limits temporal resolution, the RF network provides critical coverage where the radar is weakest. This effect would be even more pronounced for flight areas farther north of Fig.~\ref{fig:radar_traj_error} where the radar's LOS will often be obstructed by trees. This complementary performance reinforces the need for sensor fusion: the RF sensors stabilize tracking in radar-degraded regions, while the radar compensates for RF sensor dropouts and errors.

\subsection{Kalman Filter Fusion and Tracking Accuracy}

\begin{figure}
    \centering
    \includegraphics[width=1\linewidth]{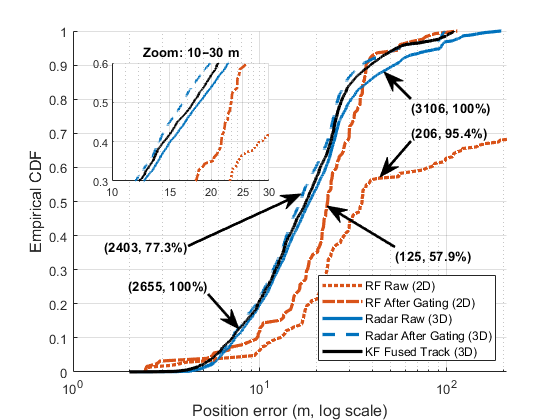}
    \caption{Empirical CDFs of position errors (log scale) for RF sensors, radar, and KF fusion. Number of available measurements and associated coverage for each scenario are also noted.}
    \label{fig:cdf}
\end{figure}

The KF integrates radar and RF measurements into a unified 3D state estimate, updating whenever a validated measurement is available and propagating the state forward when no measurements are available (coasting). This enables continuous tracking across the entire trajectory regardless of individual sensor dropouts. 

As shown in the empirical cumulative distribution functions (CDFs) of Fig.~\ref{fig:cdf}, the KF improves performance relative to the standalone sensing methods before validation, eliminating large outliers present in both the raw RF and radar measurements. After sensor validation, the radar achieves the lowest standalone mean error of 21.0 m with a standard deviation of 17.1 m, but Table~\ref{tab:position_error_stats} shows that it provides only 77.3\% coverage due to long-range degradation and track fragmentation. In contrast, the fused KF output achieves nearly identical accuracy---with a mean error of 21.9 m and a standard deviation of 17.9 m---while maintaining 100\% coverage across the full trajectory. Because the Fortem radar measurements are themselves likely the output of an onboard tracking filter, their errors are already partially smoothed; accordingly, we would expect both higher standalone radar errors and a more pronounced KF improvement if the radar were reporting raw detections rather than internally filtered estimates. The KF also produces stable predictions during short periods without measurements, with its coasted estimates exhibiting a mean error of 28.3 m and a standard deviation of 18.4 m. 

These results illustrate that the primary benefit of fusion is not a reduction in mean error, but the suppression of large outliers and the extension of tracking capability to provide increased range and full-trajectory coverage. Overall, fusion leverages the complementary strengths of the RF and radar sensors to deliver full-coverage, high-quality tracking that neither modality can achieve alone.

\section{Conclusion}

This work evaluated a multi-modal UAV tracking system that combines distributed RF sensors with a Ku-band radar, reflecting emerging ISAC-style approaches that leverage communication and sensing infrastructure together. Standalone results showed that each modality performs well under different geometry and range conditions, motivating the need for fusion. A Kalman filter with a constant-velocity motion model improved accuracy over the individual sensing methods while providing additional coverage and suppressing large errors. Use of RF sensors over a larger area is expected to further improve coverage. These findings demonstrate the value of complementary sensing for reliable UAV tracking in outdoor environments. Future work will explore incorporating additional sensing modalities into the fusion framework, applying deep-learning-based fusion methods, and using adaptive covariance tuning to improve robustness.



\bibliographystyle{IEEEtran}
\bibliography{references.bib}

\end{document}